\documentclass[11pt,fleqn]{article}
\usepackage{epsfig}
\usepackage{xspace}

\newcommand{\scaption}[1]{\parbox{1.0\textwidth}{
    \caption{\protect{\it #1}}}}
\newcommand{\itemp}{\protect\vspace*{-0.2cm}\item}

\newcommand{\psinc}[1]{\vspace{-2mm}\epsfig{clip=,#1}\vspace{-5mm}}

\newcommand{\Znull}{\ensuremath{Z^0}\xspace}

\newcommand{\SUtwo}{\ensuremath{SU_{2}}\xspace}
\newcommand{\Uone}{\ensuremath{U_{1}}\xspace}

\newcommand{\LSP}{\ensuremath{\chi_1^0}\xspace}
\newcommand{\MLSP}{\ensuremath{M_{\chi_1^0}}\xspace}
\newcommand{\tanb}{\ensuremath{\tan{\!\beta}}\xspace}
\newcommand{\thetaw}{\ensuremath{\theta_W}\xspace}
\newcommand{\MU}{\ensuremath{\mu}\xspace}
\newcommand{\Mtwo}{\ensuremath{M_2}\xspace}
\newcommand{\Mone}{\ensuremath{M_1}\xspace}
\newcommand{\Mse}{\ensuremath{M_{\tilde{e}}}\xspace}
\newcommand{\Msnu}{\ensuremath{M_{\tilde{\nu}}}\xspace}
\newcommand{\Msq}{\ensuremath{M_{\tilde{q}}}\xspace}
\newcommand{\msemsqhalf}{\ensuremath{(\Mse+\Msq)/2}\xspace}
\newcommand{\Bselsp}{\ensuremath{{\rm B}(\se\rightarrow\LSP)}\xspace}
\newcommand{\Bsqlsp}{\ensuremath{{\rm B}(\sq\rightarrow\LSP)}\xspace}

\newcommand{\se}{\ensuremath{\tilde{e}}\xspace}
\newcommand{\sq}{\ensuremath{\tilde{q}}\xspace}
\newcommand{\su}{\ensuremath{\tilde{u}}\xspace}
\newcommand{\sd}{\ensuremath{\tilde{d}}\xspace}
\newcommand{\snu}{\ensuremath{\tilde{\nu}}\xspace}

\newcommand{\PT}{\ensuremath{P_{t}}\xspace}
\newcommand{\EPZ}{\ensuremath{E - P_{z}}\xspace}

\newcommand{\PTpar}{\ensuremath{P_{t\parallel}}\xspace}
\newcommand{\PTperp}{\ensuremath{P_{t\perp}}\xspace}

\newcommand{\invpb}{\ensuremath{\mathrm{pb}^{-1}}\xspace}

\newcommand{\TeV}{\mbox{\rm ~TeV}\xspace}
\newcommand{\GeV}{\mbox{\rm ~GeV}\xspace}

\begin{document}

\begin{titlepage}

\thispagestyle{empty}
\setcounter{page}{1}

\noindent
{\tt DESY 96-082 \hfill ISSN 0418-9833\\
April 1996}
\vspace*{5.5cm}

\begin{center}
\boldmath
{\LARGE \bf A Search for Selectrons and Squarks at HERA}\\
\unboldmath 

\vspace*{1.5cm}

{\Large \bf H1 Collaboration}\\
\end{center}

\vspace*{3cm}

\begin{abstract}
\normalsize
\noindent
Data from electron--proton collisions at a center-of-mass energy of 300 GeV are
used for 
a search for selectrons and squarks within the framework of the minimal
supersymmetric model.
The  decays of selectrons and squarks 
into the lightest supersymmetric particle lead to  final states
with an electron and hadrons accompanied by large missing
energy and transverse momentum.
No signal is found and new bounds on the existence of these particles
are derived. 
At 95\% confidence level the excluded region extends to 65\GeV for
selectron and squark masses, and to 40\GeV 
for the mass of the lightest supersymmetric particle.

\end{abstract}

\end{titlepage}

{\normalsize \raggedright
\clearpage
 S.~Aid$^{14}$,                   %
 V.~Andreev$^{26}$,               %
 B.~Andrieu$^{29}$,               %
 R.-D.~Appuhn$^{12}$,             %
 M.~Arpagaus$^{37}$,              %
 A.~Babaev$^{25}$,                %
 J.~B\"ahr$^{36}$,                %
 J.~B\'an$^{18}$,                 %
 Y.~Ban$^{28}$,                   %
 P.~Baranov$^{26}$
% ...

\smallskip
% ...
$ ^{12}$ DESY, Hamburg, Germany$^a$ \\
% ...
$ ^{14}$ II. Institut f\"ur Experimentalphysik, Universit\"at Hamburg,
                                     Hamburg, Germany$^ a$  \\
% ...
$ ^{18}$ Institute of Experimental Physics, Slovak Academy of
                Sciences, Ko\v{s}ice, Slovak Republic$^ f$\\
% ...
$ ^{25}$ Institute for Theoretical and Experimental Physics,
                                                 Moscow, Russia \\
 $ ^{26}$ Lebedev Physical Institute, Moscow, Russia$^ f$ \\
% ...
$ ^{28}$ LAL, Universit\'{e} de Paris-Sud, IN2P3-CNRS,
                            Orsay, France\\
 $ ^{29}$ LPNHE, Ecole Polytechnique, IN2P3-CNRS,
                             Palaiseau, France \\
% ...
$ ^{36}$ DESY, Institut f\"ur Hochenergiephysik,
                              Zeuthen, Germany$^ a$\\
 $ ^{37}$ Institut f\"ur Teilchenphysik,
          ETH, Z\"urich, Switzerland$^ i$\\
% ...
$ ^a$ Supported by the Bundesministerium f\"ur Bildung, Wissenschaft,
        Forschung und Technologie, FRG,
        under contract numbers 6AC17P, 6AC47P, 6DO57I, 6HH17P, 6HH27I,
        6HD17I, 6HD27I, 6KI17P, 6MP17I, and 6WT87P \\
% ...
$ ^f$ Supported by the Deutsche Forschungsgemeinschaft\\
% ...
$ ^i$ Supported by the Swiss National Science Foundation\\

}
\clearpage

\section{Introduction}

Supersymmetry \cite{susy} is presently considered to be a promising
candidate for a theory beyond the Standard Model (SM) of particle
physics. In particular the Minimal Supersymmetric extension of the Standard
Model (MSSM) describes as well as the SM all experimental data, and in
addition it offers solutions for some of the questions left open by the
SM, such as the Higgs mass  or the hierarchy problem.  However,
no direct evidence for supersymmetry has yet been found.

Supersymmetry relates fermions to bosons and predicts
for each SM particle a partner  with 
spin differing by half a unit. 
So sneutrinos and selectrons, $(\snu_{eL}, \se_L),  \; \se_R$,
are scalar partners of 
neutrinos and electrons, $(\nu_{eL},e_L), e_R$, and similarly
squarks $(\su_L,\sd_L),\su_R,\sd_R$ are the partners
of up and down quarks.
Two Higgs doublets with vacuum expectation values $v_2,v_1$ are necessary 
to give masses to up-type quarks ($v_2$) and to down-type quarks and charged 
leptons ($v_1$).
The partners of the gauge bosons $W^{\pm}, Z^0,\gamma $ and the two Higgs 
doublets are called gauginos and higgsinos.
They can mix 
and form two charged mass eigenstates $\chi_{1,2}^{\pm}$ (charginos)
and four neutral mass eigenstates $\chi^0_{1,2,3,4}$ (neutralinos).

Experimentally supersymmetric particles are constrained
to be heavier than 
their SM partners and so supersymmetry must be broken. 
In the MSSM this leads to  extra mass 
parameters \Mtwo and \Mone for the \SUtwo and  \Uone gauginos.
Thus the masses of charginos and neutralinos depend on 
\Mone, \Mtwo, 
$\tanb\equiv v_2/v_1$ and the 
higgsino mass parameter \MU. 
The gaugino--sfermion couplings are the same as in the SM, but via 
mixing the chargino and neutralino--sfermion couplings also depend on these 
supersymmetric parameters.

In the MSSM it is assumed that the multiplicative quantum number $R$-parity
is conserved, where $R_P=1$ for the SM particles and $R_P=-1$
for their superpartners.
This implies that supersymmetric particles can only be produced in pairs
and that the lightest of them, which generally is assumed to be the
\LSP, is stable.  Since the \LSP is neutral and only weakly interacting
it will escape direct experimental detection.

\begin{figure}[htb]
\begin{center}
\epsfig{width=6cm,clip=,bbllx=229,bblly=539,bburx=407,bbury=700,file=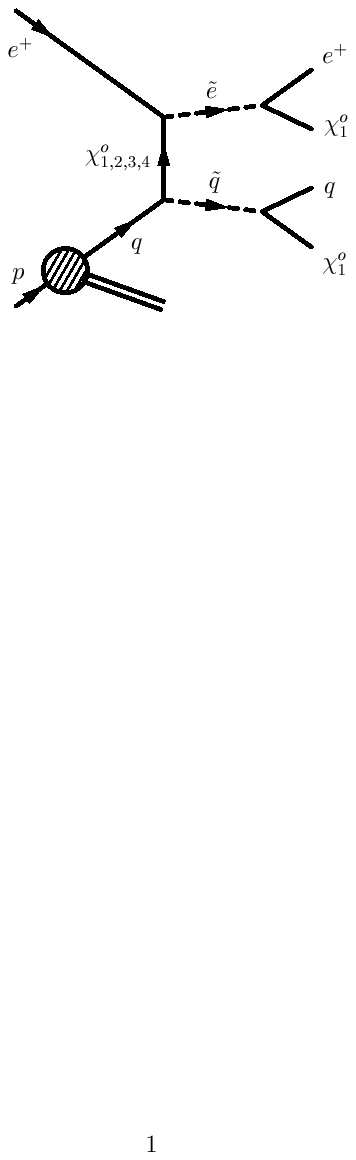}
\scaption{
  Feynman diagram for selectron - squark production via neutralino
  exchange and the subsequent decays into the lightest supersymmetric
  particle \LSP .
  \label{fig0}
  }
\end{center}
\end{figure}

At HERA the dominant MSSM process is the
production  of
a selectron and a squark via neutralino exchange 
    $e p \rightarrow \se\ \sq\ X$ as shown in Fig.~\ref{fig0}.
The \se and \sq decay into any lighter gaugino and their SM partners.
The decay involving \LSP gives an experimentally clean
signature of an electron, hadrons and missing energy and momentum.
This is the channel analyzed here, where the proper 
branching ratios for $\se \rightarrow e \LSP$  and
 $\sq \rightarrow q \LSP$ are taken into account.

In this paper it is assumed that
the masses of $\se_L$ and $\se_R$ are
identical. The same holds for the \sq\ states and mass
degeneracy is also assumed for the superpartners of the four lightest quarks.
In order to reduce the number of free MSSM parameters  it is
assumed that \Mone and \Mtwo are related via the weak mixing angle
\thetaw,  $\Mone = 5/3 ~\Mtwo~ \tan^2{\thetaw}$, 
as  suggested by Grand Unified Theories (GUT's) \cite{susy}. 
No other GUT
relations are used. The parameter set given above specifies completely all 
couplings and 
masses involved.

At $e^+e^-$ experiments masses below 45 \GeV for selectrons and for
squarks have been excluded by direct searches \cite{lep1} and no attempt
was made here to evaluate this excluded region.  A search for selectron
production at 130 and 136 GeV \protect\cite{aleph96,l3_96} yielded a lower
limit of $\Mse=56 \GeV$ for $\MLSP<35 \GeV$, %
but however, with different assumptions for the $\se_R , \, \se_L$ mass
splitting.

At the $Z^0$ resonance direct searches for charginos and neutralinos
\cite{l3_95} restrict the MSSM parameter space mainly for higgsino--like
\LSP and for \LSP masses up to 20 GeV. These results are independent of
\Mse and \Msnu.
Above the $Z^0$ resonance recent searches have extended these limits
considerably \protect\cite{aleph96,opal96,l3_96}. 
Here also \se and \snu exchange contributes and, due to destructive
interference, the constraints are most stringent for large \Mse and \Msnu.

The experiments at the Tevatron $p\bar p$ collider have obtained strong bounds
on \sq\ masses. These limits however either depend on the assumption of a light
\LSP
\cite{cdf92} or they are only valid for small
gluino masses \cite{cdf96}.  
The latter bounds can only be related to the results
obtained here if one assumes additional GUT relations between the gluino
mass and \Mtwo, \Mone.

While this search deals with supersymmetric models with conserved $R_P$,
the  production of \sq in models with broken $R_P$ has been 
considered in \cite{h1rparity}.

\section{Detector description}
 
A detailed description of the H1 apparatus can be found elsewhere \cite{H1NIM}.
The following briefly describes the components of the detector relevant
to this analysis, which makes use mainly of the calorimeters.

The liquid Argon (LAr) calorimeter \cite{LARC} extends with full
azimuthal coverage over the polar angular range
$4^\circ < \theta <  153^\circ$, 
where $\theta$ is defined with respect to the proton beam direction
($+z$ axis). 
The calorimeter  is highly segmented and consists of an electromagnetic 
section with
lead absorbers, corresponding to a depth of between 20 and 30 radiation
lengths, and a hadronic section with steel absorbers.
The total depth of the LAr~calorimeter varies between 4.5 and 8
hadronic interaction lengths. 
Single particles are measured in the LAr~calorimeter with energy
resolutions of  $\sigma(E)/E\approx 0.11/\sqrt{E/\mbox{GeV}}\oplus 0.01$
for electrons
and $\sigma(E)/E\approx 0.5/\sqrt{E/\mbox{GeV}}\oplus 0.02$
for charged pions.
The electromagnetic and hadronic energy scales
are known to 2\% and 5\% respectively.

The calorimeter is surrounded by a superconducting solenoid providing a
uniform magnetic field of $1.15$ T parallel to the beam axis in the
tracking region.  A large instrumented iron structure surrounds the
solenoid to serve as a return yoke for the magnetic flux and as an
additional calorimeter of 4.5 hadronic interaction lengths to absorb
tails of hadronic showers.

The backward region of the detector ($151^\circ < \theta < 177^\circ $)
was covered in 1994 by an electromagnetic
lead--scintillator calorimeter, which was replaced in 1995  by a new 
lead--scintillating fibre
calorimeter \cite{spacal} with improved containment and angular
coverage.
A plug calorimeter covers the angular region $0.75^\circ < \theta <
3.5^\circ$ around the forward beam pipe.

Charged particle tracks are measured in two concentric jet drift
chambers and a forward tracking detector covering the polar angular range 
$ 7^\circ < \theta < 165^\circ$.
In this analysis the tracking detectors are used for the
determination of the event vertex only.

\section{Data selection and analysis}

The event sample corresponds to an integrated luminosity of $6.38 \pm
0.13 \; \invpb$ and was collected by the H1 experiment in 1994 and 1995
during collisions of 27.5 GeV positrons\footnote{In this paper the
generic name ``electron'' is used also for positrons.}
with 820 GeV protons.

First some event quantities which are used in the analysis are
introduced.  The quantity \EPZ is defined as $\sum_i E_i (1 -
\cos\theta_i)$ where the sum runs over all LAr and instrumented iron
calorimeter energy deposits. The backward and plug calorimeters are used
only as a veto. The (2 dimensional) transverse momentum vector of a
single energy deposition is $\vec P_{t,i} = E_i \sin \theta_i (\cos
\phi_i,\sin \phi_i)$ with polar angle $\theta_i$ and azimuthal angle
$\phi_i$.  The total transverse momentum vector $\vec P_t$ is the vector
sum of all $\vec P_{t,i}$ in the LAr and iron calorimeter. The variable
\PTpar (\PTperp) is defined as the absolute value of the component of
$\vec P_t$ which is parallel (perpendicular) to the transverse component
of the identified scattered electron.

\noindent
The following selection criteria are used:
\begin{enumerate}
\itemp Events must be accepted by a LAr calorimeter trigger for
  electrons which requires a local energy deposit of more than 8 GeV in
  a small region of the electromagnetic LAr calorimeter.\nolinebreak
\itemp A vertex has to be found from measured charged tracks
  to be within 35 cm of the nominal interaction point.
\itemp An electron must be identified by its shower shape in the
  LAr calorimeter, it must have an energy 
  greater than 10 GeV, and its transverse momentum 
  has to exceed  8 GeV.
  The polar angle $\theta_e$  must be in the range
  $10 ^\circ < \theta_e < 135 ^\circ$ and 
  the electron must be isolated in
  a cone of pseudorapidity and azimuthal angle of radius 0.5.
\itemp The energy deposited in the backward calorimeter
  has to be below $5\GeV$.
\itemp The transverse momentum measured in the plug
  calorimeter must be below $2 \GeV$.
\itemp Filters against cosmic and beam halo muons are applied.
\itemp The quantity \EPZ must be below 
  $40 \GeV$. 
\itemp Either \label{altcut}
  \begin{enumerate}
  \item the polar angle of the ``current quark'' must fulfill $\theta_q >5
    ^\circ $.
    Assuming standard deep inelastic scattering kinematics, $\theta_q$
    can be calculated in the quark parton model from the scattered
    electron and the electron beam energy.
    To compensate for an undetected radiative photon in the initial
    state, the electron beam energy is reduced by the energy of this
    photon which is determined using the measured total \EPZ.
  \item
    $ \PTperp >3\GeV$  and
  \item $\PTpar > 2.5 \GeV + (\EPZ)/2$
  \end{enumerate}
  or
  \begin{enumerate}
    \addtocounter{enumii}{3}
  \item $\PTperp > 7 \GeV$  and
  \item $\PTpar > 3 \GeV$
  \end{enumerate}
\end{enumerate}
The asymmetry of the HERA beam energies leads to a boost in the proton
beam direction of the produced heavy \se and \sq and their decay
products. Therefore, in contrast to the majority of SM events,
no energy deposition is expected in the backward region of the detector
(cuts 3 and 4).  Energy losses in the forward beam pipe can create large
values of $P_t$. Such events are vetoed by cut 5.

\begin{figure}[htbp]
\vspace*{5mm}
\begin{center}
\epsfig{width=13cm,clip=,file=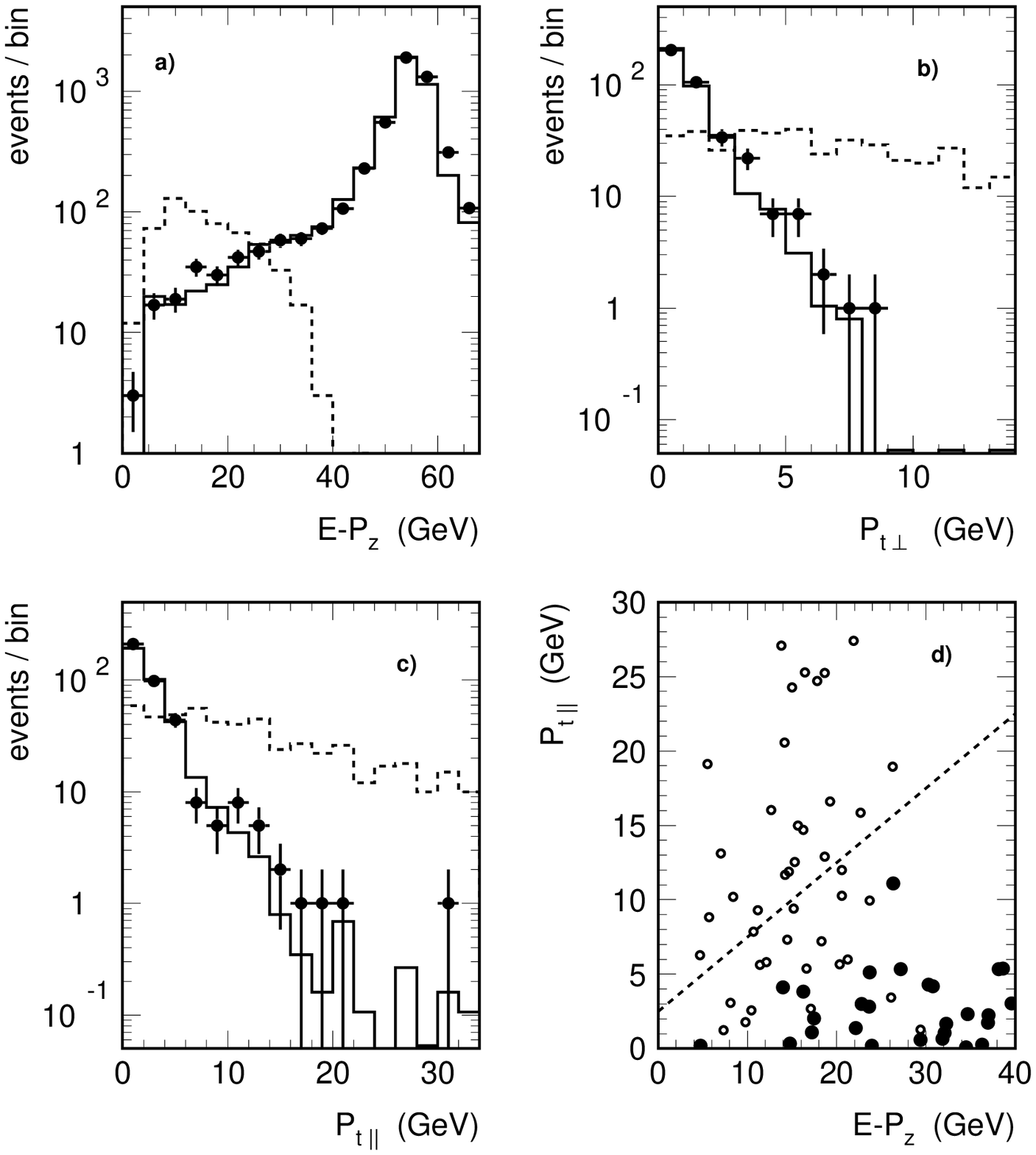}
\scaption{
  Comparison of data (points) with the
  expectation for all contributing SM processes (solid histogram). To
  indicate the signal region,  simulated \se\sq events
  are shown with arbitrary normalization by the dashed histogram for
  $\Mse=\Msq=65\GeV$ and   \MLSP between 35 and 55 GeV.
  Figure a) shows the \EPZ distribution after basic electron selection
  cuts 1--6. Figures b) and c) show the \PT component 
  perpendicular (\PTperp) and parallel
  (\PTpar) to the electron direction after applying an additional cut of 
  $\EPZ < 40 \GeV$. 
  Figure d) shows data (solid dots) and simulated \se\sq events
  (open circles) after applying cuts 1--7, 8a and 8b (see text).  The
  combined cut on $\PTpar > 2.5 \GeV + (\EPZ)/2$ is indicated
  by the dashed line.
  \label{fig1}
  }
\end{center}
\end{figure}

Because of momentum conservation,
 the value $\EPZ=2\times27.5\GeV=55\GeV$ of the
initial state beam particles is expected also for the measured
final state of most SM processes.
Supersymmetric events would have a much smaller value of \EPZ
due to the undetected \LSP's. 
This is demonstrated by the dashed histogram in
Fig.~\ref{fig1}a which shows a superposition of several hundred simulated
\se\sq events with $\Mse = \Msq = 65 
\GeV$ and various values of \MLSP.

The event generator HERASUSY \cite{herasusy}, which is used for this 
simulation, is based on the
cross section for all four neutralino exchange diagrams and their 
interference \cite{litsesq}. All
couplings and masses involved are determined using a modified
version of the ISASUSY program \cite{isasusy}.
The PYTHIA program \cite{pythia}  is used for parton showers, string
fragmentation and particle decays and
a detailed simulation of the H1 detector response is included.

The events selected by cuts 1 -- 6 are shown as solid points with
statistical errors in
Fig.~\ref{fig1}a. They 
exhibit a clear peak at $\EPZ = 55 \GeV$ and a long tail
towards lower values.  
This distribution is well described by the solid
histogram which represents a simulation of the contributing SM processes,
i.e.
neutral current and charged current deep inelastic scattering events 
simulated using the DJANGO program \cite{django}
and photoproduction events simulated using the PYTHIA
program \cite{pythia}.
The tail towards low \EPZ is due to particles lost in the $-z$
direction
and is populated mainly by neutral current events with a photon
radiated off the initial state electron.  
Contributing at very low \EPZ are also photoproduction processes, in
which the scattered electron is not detected and a hard photon is
misidentified as an electron in the event selection.

After cut 7 motivated by the HERASUSY simulation 
Fig.~\ref{fig1}b and c show the distributions of \PTperp and \PTpar
for the remaining 384 events.
$367\pm19$ events are expected from all background sources.
The resolution for \PTperp is much better than for \PTpar since more
energy is deposited collinear to the electron in SM processes.  The
extreme tails of the $P_t$ distributions are populated by radiative
charged current 
events with a photon misidentified as an electron.  In all three
distributions no obvious excess over expectation is observed.

Cut 8a on the jet angle predicted by the measured
electron is employed 
in order to reduce the neutral current background of events with jets lost
in the beam pipe. 
Also a minimum value of \PTperp
is required (cut 8b).
Cut 8c on the correlation of \PTpar and \EPZ is illustrated in 
Fig.~\ref{fig1}d which shows the data after 
cuts 1--7 and 8a together with a
sample of  simulated \se\sq events.
Cut 8c leads to a considerable  loss for \se\sq events with small
values of \PTpar. A large fraction of these events are recovered by
an alternative choice of cuts requesting mainly a large value of \PTperp
(cuts 8d and 8e).

After all cuts no events are found in a simulation of neutral current
(photoproduction) processes in which there were 3 (0.63) times the statistics of
the data. A background of $0.6 \pm 0.2$ radiative charged current events
is expected. 
No candidate event remains in the data.

\section{Results}

Having observed no signal, exclusion limits for \se\sq production are
derived.

The selection efficiency is determined using events generated as
described above with the HERASUSY program for different values of \Mse,
\Msq and \MLSP.
Other MSSM parameters do not influence the final state kinematics
significantly. In order to interpolate between the simulated values for 
\Mse, \Msq and \MLSP an empirical function $\epsilon(\mathcal{P})$ 
is fitted to the simulated efficiency values.
Here $\mathcal{P}$ is defined via 
$$\mathcal{P}^2 \equiv    \frac{\Mse^2 - \MLSP^2}{2\;\Mse}
                 \;\times\;\frac{\Msq^2 - \MLSP^2}{2\;\Msq}.$$
It characterizes the transverse momenta  of electron and quark after
the 2-body decays $\se \rightarrow e \LSP$ and $\sq \rightarrow q \LSP$.
The efficiency $\epsilon(\mathcal{P})$ reaches a plateau of 56\% for
 $\mathcal{P}>25\GeV$. For smaller values of $\mathcal{P}$ any signal in
the detector becomes less and less significant and the efficiency drops to 50\%,
18\% and 0\% for values of $\mathcal{P} =$ 20, 10 and 5\GeV
respectively.

Systematic errors of this analysis originate in the efficiency
parameterization, the electromagnetic and hadronic calorimeter energy
scales, the parton density functions and the integrated luminosity.  The
total systematic uncertainty amounts to 4.3\% in the number
of expected supersymmetric events.

\begin{figure}[p]
\begin{center}
\vspace{-20mm}
\psinc{width=12cm,file=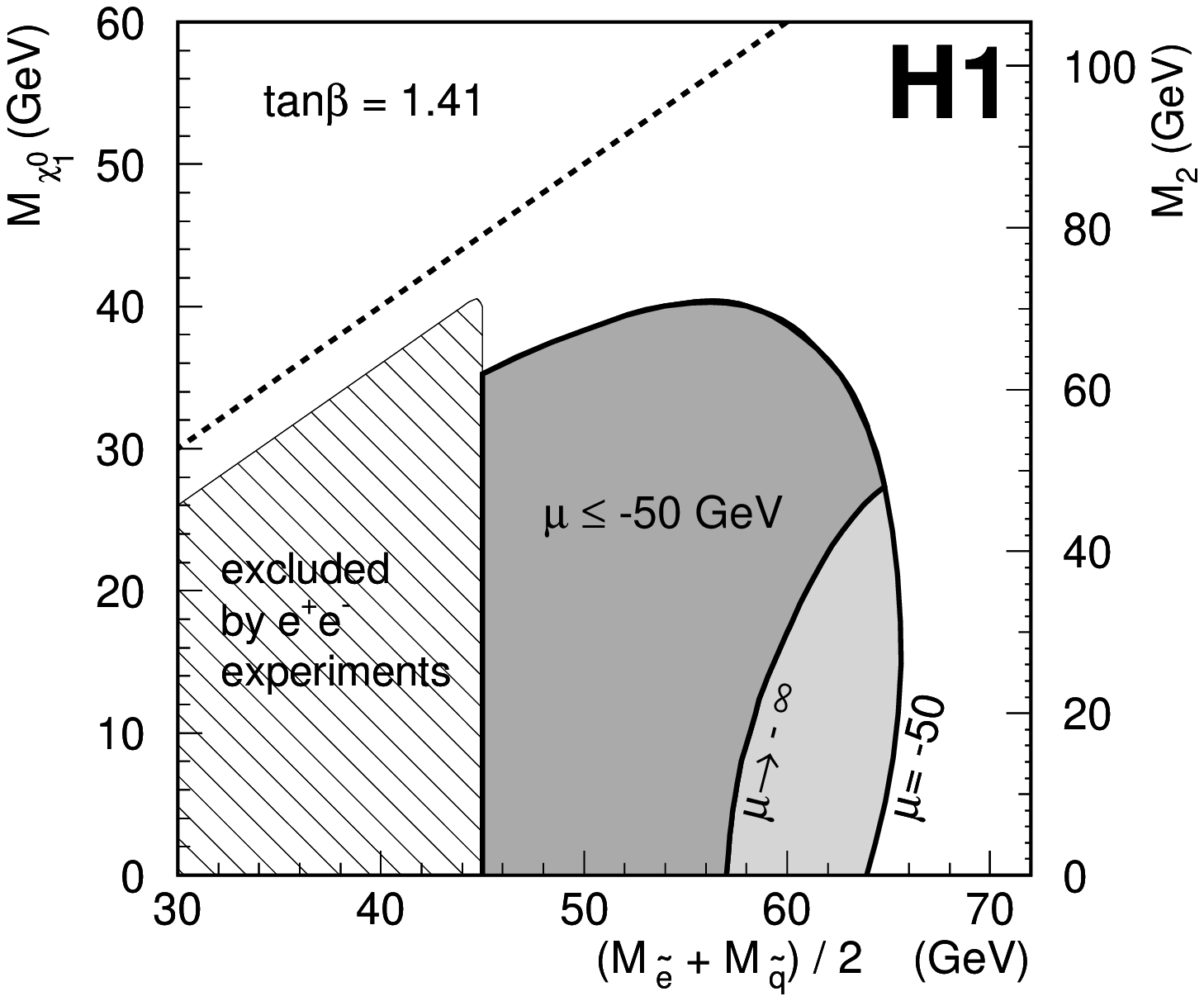}
\scaption{
  Excluded regions at 95\% confidence level in the plane of \MLSP versus
  half the sum of \Mse and \Msq  for $\tanb =  1.41$.
  The dark and light shaded areas are excluded for $\MU = -50\GeV$.
  The dark shaded area is excluded for all $\MU \leq -50\GeV$. 
  The difference between both is due to decays
  into other gauginos than the \LSP. 
  The  diagonal line corresponds to $M_{\se,\sq}=\MLSP$.
  The hatched area is excluded by direct searches for \se and \sq pair
  production in $Z^0$ decays \protect\cite{lep1}. 
  The vertical axis is
  also labeled in terms of \Mtwo (see text).
  \label{mlspmse}
\protect\vspace{-10mm}
  }
\psinc{width=12cm,file=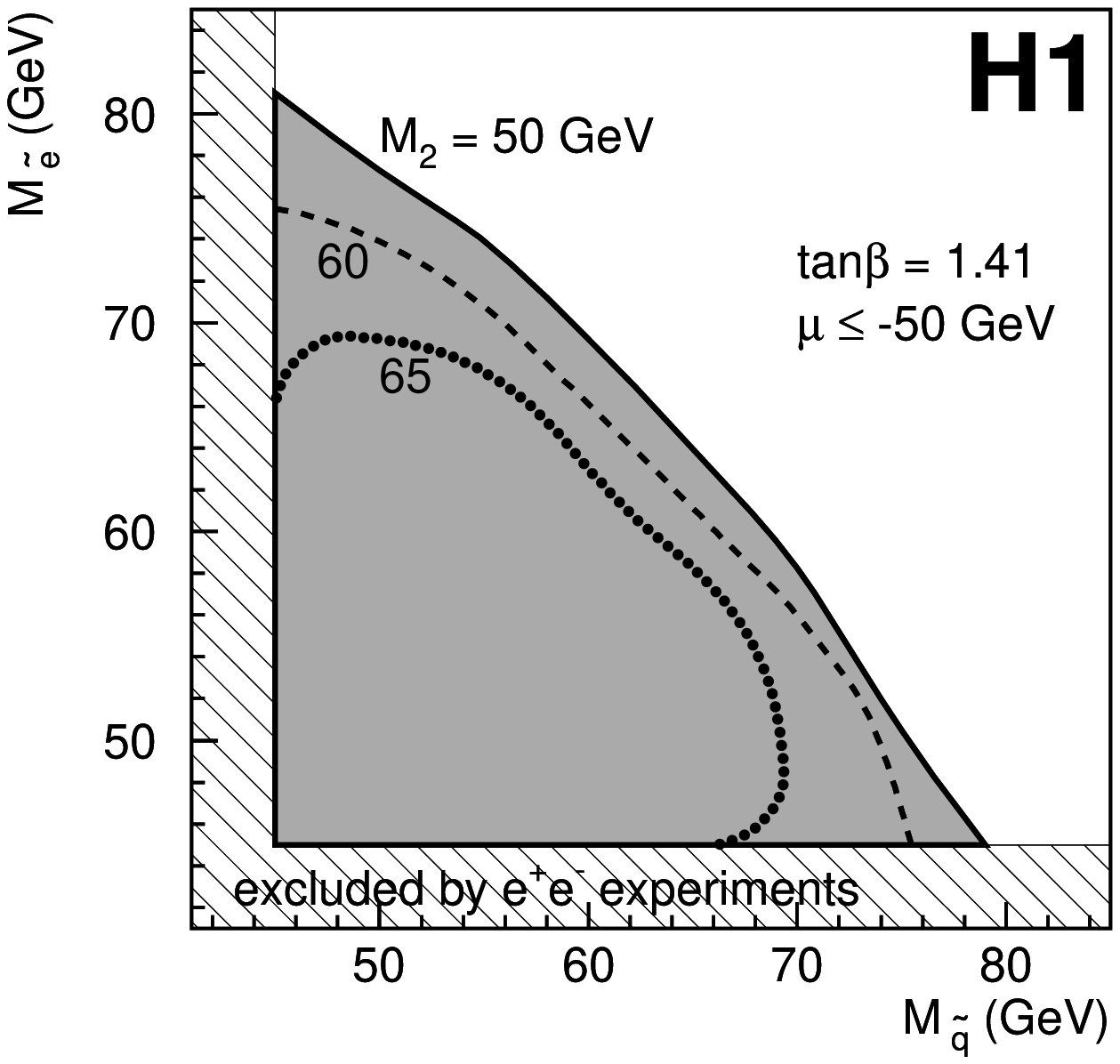}
\scaption{
  Exclusion limits at 95\%   confidence level 
  in the \Mse versus \Msq plane for different values of \Mtwo
  and $\mu \leq -50 \GeV$, $\tanb = 1.41$.
  \Mse and \Msq masses below 45\GeV are excluded by searches in \Znull decays.
  \protect\cite{lep1}. 
  \label{msemsq}
  }
\end{center}
\end{figure}

\begin{figure}[htb]
\begin{center}
\vspace{-10mm}
\psinc{width=12cm,file=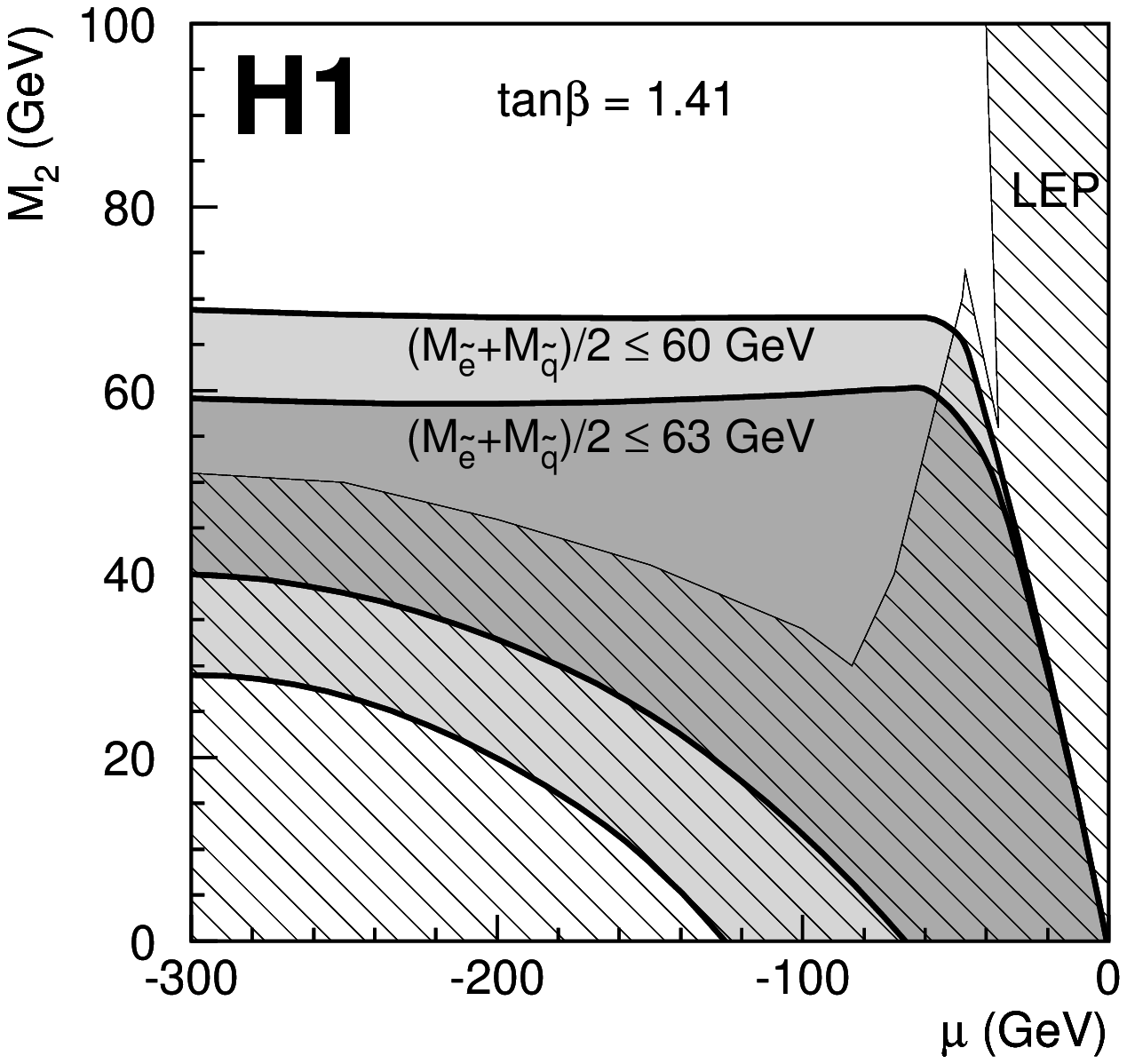}
\scaption{
  Domains  excluded at 95\% confidence   level
  in the \Mtwo versus \MU plane from the search for \se\sq production for
  $\tanb=1.41$. The dark shaded area corresponds to
  $\msemsqhalf \leq 63\GeV$, the additional light shaded area to
  $\msemsqhalf \leq 60\GeV$. 
  The hatched areas show the most restrictive bounds 
  obtained from $e^+e^-$ collisions at LEP at center-of-mass energies around the
  \Znull mass and at 130 and 136\GeV  \protect\cite{aleph96}, see text.
\label{m2mu}
}
\end{center}
\end{figure}

\begin{figure}[htb]
\begin{center}
\psinc{width=12cm,file=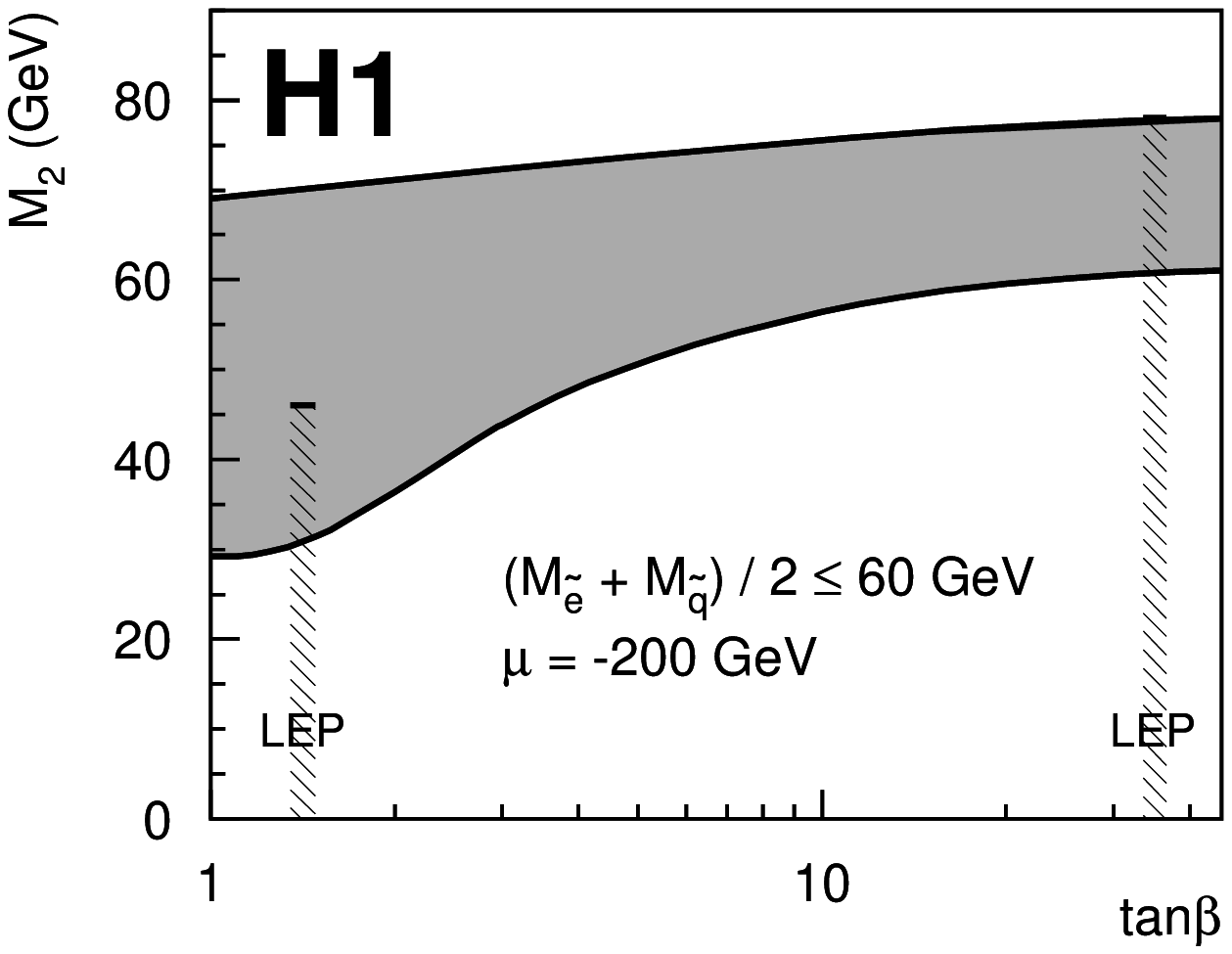}
\scaption{
  Excluded region at 95\%  confidence level for \Mtwo  as a
  function of \tanb for 
  \mbox{$\msemsqhalf \leq 60 \GeV$ and $\mu = -200 \GeV$} (shaded area).
  The two
  symbols show the upper limits on \Mtwo obtained for $\tanb=1.41$ and
  $35$ from $e^+e^-$
  collisions at LEP for center-of-mass energies 
  of 130 and 136 \GeV \protect\cite{aleph96}.
\label{m2tanb}
}
\end{center}
\end{figure}

In order to determine excluded regions in the MSSM parameter space,
a systematic scan 
is performed for values of $\Mse\geq45\GeV$, $\Msq\geq45\GeV$,
 $-1\TeV\leq\MU\leq1\TeV $, $\Mtwo\geq 0\GeV$ and $1 \leq \tanb \leq 50$.
For each chosen set of these parameters the cross section, branching
ratios and efficiency are evaluated yielding an expectation for the
number of observed signal events. 
A parameter set is excluded when this number exceeds the
95\% confidence level upper limit for 0 observed events taking into
account the 4.3\% systematic uncertainty mentioned above. 
Details of this statistical procedure are described in \cite{h1poisson}.
The resulting  excluded regions in the MSSM parameter space are
presented  in the planes   
\begin{itemize}
\itemp  \MLSP versus $\msemsqhalf$  for constant \MU and $\tanb=1.41$
             (Fig.~\ref{mlspmse}) 
\itemp  \Mse versus \Msq for constant \Mtwo, \MU and $\tanb=1.41$
             (Fig.~\ref{msemsq})
\itemp  \Mtwo versus \MU for constant $\msemsqhalf$
             and $\tanb=1.41$ (Fig.~\ref{m2mu})
\itemp  \Mtwo versus \tanb for constant $\msemsqhalf$ and \MU
             (Fig.~\ref{m2tanb})
\end{itemize}
The choice of $\tanb=1.41$ is taken as typical of a low value for \tanb, and
allows direct comparison with recent results obtained at LEP.
The dependence on \tanb is shown in Fig. \ref{m2tanb}.

The  cross section for  \se\sq production depends to first approximation
on the \LSP  mass  and couplings, and on  the energy
threshold for the hard process $\Mse+\Msq$. 
The  neutralino and chargino masses scale roughly with \Mtwo for
$\MU \ll 0$.
For the excluded parameter regions
the branching ratios \Bselsp and \Bsqlsp are close to
one because all other gauginos are heavier than \Mse or \Msq. 
Only at very small \MU and \Mtwo other decays 
are possible, which have not been searched for in this paper.

In Fig.~\ref{mlspmse} the plane \MLSP versus \msemsqhalf is shown 
for $\tanb=1.41$. 
The dark and light shaded areas are excluded for $\MU = -50\GeV$.
For smaller values of \MU ($\MU \rightarrow - \infty$) 
and for small \Mtwo the competing decays into
other gauginos lead to reduced limits for \msemsqhalf. 
The dark shaded area is excluded for all $\MU \leq -50\GeV$.
In the range $\MU\leq-50\GeV$, and for this fixed value of \tanb, the \LSP
mass depends on \Mtwo only, which is indicated by the second vertical
scale on the right side of the figure.
At large \MLSP the mass difference between \se, \sq and \LSP becomes
small and the efficiency drops correspondingly.
Also the cross section is reduced due to the higher propagator
mass. Both effects lead to reduced limits for \msemsqhalf.
As seen from the distance between the shaded area and the diagonal line
for $\MLSP=M_{\se,\sq}$ this search is sensitive down to mass differences of
$\approx 10\GeV$. 
The excluded mass range extend to  65\GeV for \msemsqhalf and to
40\GeV for \MLSP. In particular the region $\msemsqhalf\leq63\GeV$ is
excluded for $\MLSP\leq35\GeV$.
Also shown are the limits obtained from direct searches for \se and \sq
pair production in $Z^0$ decays \cite{lep1}.

Fig.~\ref{msemsq} shows the excluded regions in \Mse versus \Msq for three
different values of \Mtwo.
For a large range of \Mse and \Msq the limits are approximately
a function of only the sum $\Mse+\Msq$, as used in
Fig. \ref{mlspmse}. This approximation is valid for large mass
differences between \se and \LSP and also between \sq and \LSP,
i.e. $M_{\se,\sq}-\MLSP \geq 20\GeV$.
The limits on \Mse and \Msq decrease when these  differences
become too small.
For $\Mtwo \approx 50\GeV$ the highest mass bounds are reached, namely
 $\Mse=77\GeV$ at $\Msq=50\GeV$ and $\Msq=75\GeV$ at $\Mse=50\GeV$.

In Fig.~\ref{m2mu} 
the exclusion limits are shown  in the \Mtwo versus \MU\ plane for
$\tanb=1.41$ 
and the mass ranges $\msemsqhalf \leq 60 \GeV$ and $\leq 63 \GeV$.
For $\MU \ll 0$ the \LSP is dominated by its photino
component. Consequently the coupling of \se and  \sq to the \LSP are of
electromagnetic strength and allow for a sizeable cross section. 
When \MU approaches $0$ the \LSP becomes higgsino--like and the couplings
and cross sections become very small. This defines the limit of the excluded
parameter space for \MU close to $0$.
For $\MU \ll 0$ the range excluded for \Mtwo becomes independent of \MU.
The region of $\MU \ll 0$ and small \Mtwo  is  not excluded
because of decays into charginos and
corresponds to the light shaded area in
Fig. \ref{mlspmse}.

Also displayed  as hatched areas in Fig.~\ref{m2mu} are the most
stringent 
bounds from searches for charginos and neutralinos at
LEP at center-of-mass energies around the \Znull mass and at 130 and 136 GeV
\protect\cite{aleph96,opal96}.
These limits are valid for $M_{\se,\snu}\ge500\GeV$. For smaller masses
the bounds are less restrictive. 
For $\MU \ll 0$ the search presented here extends to considerably larger
values for \Mtwo than the LEP results.  For $\MU>0$ however the LEP
results are more 
restrictive. Therefore only values for $\MU\leq0$ are
displayed.

In Fig.~\ref{m2tanb} the dependence of the \Mtwo limits on \tanb is
displayed for  $\msemsqhalf\leq 60 \GeV$ and $\mu = -200 \GeV$.
The limit on \Mtwo rises to 80 GeV for large values of
\tanb. 
The two symbols at $\tanb = 1.41 $ and $35$ 
indicated the bounds from LEP as discussed above \protect\cite{aleph96}.
The range of \Mtwo excluded in addition to those of LEP is largest for
small  values of \tanb. 

\section{Conclusion}

Within the MSSM with conserved $R$-parity we have searched for 
\se\sq pairs and their dominant decay modes into the lightest
supersymmetric particle \LSP. No signal was found
and rejection limits at 95\% confidence level are derived for 
the parameters \Mse, \Msq, \Mtwo, \MU and \tanb.
The cross section, the branching ratios and the efficiency 
depend mainly on the sum $\Mse+\Msq$ and on the \LSP mass and
couplings to fermions. The search is therefore most sensitive for a \LSP
with a large photino or $\tilde{Z}$ component.
For $\tanb=1.41$ masses up to 
 $\msemsqhalf=65\GeV$ and up to $\MLSP=40\GeV$ are probed.
The dependence on \tanb is weak and mass limits improve 
towards large \tanb. 
The  parameter region\\

\centerline{$\msemsqhalf \leq 63\GeV$ for $\MLSP\leq 35\GeV$ and $\MU=-50\GeV$}

\vspace{1mm}
\noindent
is excluded. This limit applies as long as
 $M_{\se,\sq}-\MLSP\geq20\GeV$ and decreases slightly when this
difference becomes smaller. 
The result becomes independent of \MU for $\MU\leq-50\GeV $.
At small values of \Mtwo the sensitivity is reduced due to decays into
charginos.

For a photino--like \LSP and small values of \tanb these results extend
considerably  the limits obtained by the searches for supersymmetry at
LEP 1 and also those obtained while LEP was running at center of mass
energies of 130 and 136 GeV. The limits, in particular for squarks, are
independent of the gluino mass.

\section*{Acknowledgments}
We are grateful to the HERA machine group whose outstanding efforts
made this experiment possible. We appreciate the immense 
effort of the
engineers and technicians who constructed and maintained the detector.
We thank the funding agencies for their financial support of the
experiment. We wish to thank the DESY directorate for the support
and hospitality extended to the non-DESY members of the collaboration.

\end{document}